\documentclass{article}
\usepackage{spconf,graphicx,mathtools,array,float,multirow}
\newcolumntype{P}[1]{>{\centering\arraybackslash}p{#1}}
\usepackage[utf8]{inputenc}
\usepackage[english]{babel}
\usepackage{color}
\usepackage{framed}
\usepackage{cite}
\usepackage{amsmath,amssymb,amsfonts}
\usepackage{graphicx}
\usepackage{upgreek}
\usepackage{textcomp}
\usepackage{cite}
\usepackage{bbold}
\usepackage{ifpdf}
\usepackage{times}
\usepackage{layout}
\usepackage{float}
\usepackage{afterpage}
\usepackage{amstext}
\usepackage{amssymb}
\usepackage{latexsym}
\usepackage{color}
\usepackage{kantlipsum}    

\usepackage{graphicx}
\usepackage{amsmath}
\usepackage{amsthm}
\usepackage{graphicx}

\usepackage{url}
\usepackage[font=scriptsize,caption=false,labelsep=space]{subfig}

\usepackage{booktabs}
\usepackage{multicol}
\usepackage{lipsum}

\usepackage{enumitem}
\usepackage[switch]{lineno}
\usepackage[named]{algo}
\usepackage[noend]{algpseudocode}
\usepackage{algorithm}

\usepackage{seqsplit}

\usepackage{amsmath}
\usepackage{amsthm}
\usepackage{dsfont}
\usepackage{thmtools}
\usepackage{amssymb}
\usepackage[dvipsnames]{xcolor}

\def\bOmega{\boldsymbol{\Omega}}

\def\mba{\mathbf{a}}

\def\mbe{\mathbf{e}}

\def\mbn{\mathbf{n}}

\def\mbp{\mathbf{p}}

\def\mbs{\mathbf{s}}

\def\mbv{\mathbf{v}}

\def\mby{\mathbf{y}}

\def\mbA{\mathbf{A}}

\def\mbC{\mathbf{C}}

\def\mbG{\mathbf{G}}

\def\mbI{\mathbf{I}}
\def\mbJ{\mathbf{J}}
\def\mbK{\mathbf{K}}

\def\mbQ{\mathbf{Q}}
\def\mbR{\mathbf{R}}
\def\mbS{\mathbf{S}}

\def\calH{\mathcal{H}}

\def\bone{\boldsymbol{1}}

\newcommand{\complexC}[1]{\mathds{C}^{#1}}
\newcommand{\expecE}[1]{\mathds{E}\left\{{#1}\right\}}

\newcommand{\probP}[1]{\mathds{P}\mathrm{r}\left\{{#1}\right\}}

\newcommand{\realR}[1]{\mathds{R}^{#1}}

\def\Tr#1{\mathrm{Tr}\left(#1\right)}
\def\vec#1{\mathrm{vec}\left(#1\right)}

\def\Diag#1{\mathrm{Diag}\left(#1\right)}

\def\T{\top}
\def\H{\mathrm{H}}
\def\j{\mathrm{j}}

\newcommand*{\rom}[1]{\expandafter\@slowromancap\romannumeral #1@}

\usepackage{tikz}
\usetikzlibrary{shadows.blur}

\usetikzlibrary{arrows,shadows} 
\usepackage[underline=true,rounded corners=false]{pgf-umlsd}
\title{Space-Time Adaptive Processing for radars in Connected and Automated Vehicular Platoons}
\name{Zahra Esmaeilbeig$^\star$, Kumar Vijay Mishra$^\dagger$, and Mojtaba Soltanalian$^\star$ 
\thanks{This work was sponsored in part by the National Science Foundation Grant  ECCS-1809225, and in part by the Army Research Office, accomplished under Grant Number W911NF-22-1-0263. The views and conclusions contained in this document are those of the authors and should not be interpreted as representing the official policies, either expressed or implied, of the Army Research Office or the U.S. Government. The U.S. Government is authorized to reproduce and distribute reprints for
Government purposes notwithstanding any copyright notation herein.}}
\address{$^{\star}$ECE Department, University of Illinois Chicago, USA\\
 $^{\dagger}$United States DEVCOM Army Research Laboratory, USA\vspace{-10pt}}
\ninept
\begin{document}
\maketitle
\begin{abstract}
In this study,  we develop a holistic framework for space-time adaptive processing (STAP) in connected and automated vehicle (CAV) radar systems. We investigate a CAV system consisting of multiple vehicles that transmit frequency-modulated continuous-waveforms (FMCW), thereby functioning as a multistatic radar. Direct application of STAP in a network of radar systems such as in a CAV  may lead to excess interference. We  exploit time division  multiplexing (TDM) to perform  transmitter scheduling over  FMCW  pulses to  achieve high detection performance. The TDM design problem is  formulated as a  quadratic assignment problem which is tackled by power method-like iterations and  applying the  Hungarian algorithm for  linear assignment in each iteration. Numerical experiments confirm that the optimized TDM is successful in enhancing the target detection performance.
\end{abstract}
\vspace{-6pt}
\section{Introduction}
Recent innovations in connected and automated vehicle (CAV) systems provide new opportunities to enable better sensing of the environment. CAVs offer advanced collision avoidance through onboard sensor technologies like radar, cameras, and lidar~\cite{sarker2020}. Platooning in the  CAV systems is  when vehicles travel in groups with very
short inter-vehicle spacing.
The connectivity among the  vehicles in a platoon enables a CAV to receive information from surrounding vehicles and infrastructure~\cite{wang2023anomaly}. In this regard, a network of automotive radar systems assisting each other in sensing the environment is  enabled. 
Whilst a single vehicle radar may suffer from
obstruction, fading, or  lack of radial 
velocity component of the target with respect to the  radar, it is unlikely that this will be the
case with multi-vehicle  systems with  different transmitter-target-receiver paths, also known as  networked radar in the literature~\cite{Griffiths, zhangjoint2022}.

In this study, we aim to leverage the cooperative capabilities of transportation systems---specifically, vehicle-to-vehicle (V2V) and vehicle-to-infrastructure (V2I) communications---to develop a distributed space-time adaptive processing (STAP) scheme for radars in a CAV. Each vehicle is assumed to travel at a speed and in a direction that can be different from those of the other vehicles. However, the  properties of the environment,  positioning and velocity information of the vehicles in the platoon are assumed to be accessible at each vehicle through V2V communication, therefore making a cooperative  STAP scheme  feasible.  We formulate the cooperative STAP as a decentralized multistatic target  detection problem~\cite{goodman2007optimum}.

 Direct application of STAP can reduce  target  detection performance because of high sidelobes and susceptibility to interference~\cite{radarsignaldesign2022,wang2020stap,bose2021mutual,tang2016joint,tang2020polyphase,bose2022waveform,esmaeilbeig2023mutual}. On the transmitter side, the interference can be addressed by transmitting well-designed
radar signals that are nearly orthogonal to each other in the spectral or temporal domains~\cite{xu2023automotive}. In order to  enhance  robustness against interference, we  propose a time division  multiplexing(TDM)  framework in which among all the transmitters in the CAV, only one is scheduled to transmit during each pulse. In other  words, we assume that  frequency-modulated continuous-waveforms (FMCW) signals  with similar properties  but with active or silent chirps are generated.   
We consider an extended version of the transmitter scheduling framework proposed by~\cite{wang2020stap} and demonstrate that the  transmitter scheduling problem is a quadratic assignment problem (QAP) and address it  by means of  power method-like  iterations. At each iteration the  problem is  boiled down to a  linear assignment problem which is  efficiently  solved by the Hungarian algorithm~\cite{kuhn1955hungarian}.\nocite{bose2017non}

The rest of this paper is organized as follows. In the next section, we introduce the system model for the cooperative STAP in CAVs. In Section~\ref{sec:TDM}, we formulate the  design problem for obtaining the optimal TDM in our  system. Section~\ref{sec:method} presents our approach based on the  Hungarian algorithm to optimize the TDM matrix. We evaluate our methods via numerical experiments in Section~\ref{sec_5} and conclude the paper in Section~\ref{sec:conclusion}.

\emph{Notation:} Throughout this paper, we use bold lowercase and bold uppercase letters for vectors and matrices, respectively. The $(m,n)$-th element of the matrix $\mbA$ is $\mbA_{mn}$. The sets of complex and real numbers are $\mathbb{C}$ and $\mathbb{R}$, respectively;  $(\cdot)^{\top}$, $(\cdot)^{\ast}$ and $(\cdot)^{\mathrm{H}}$ are the vector/matrix transpose, conjugate, and Hermitian transpose, respectively.
The trace of a matrix is  $\operatorname{Tr}(.)$; the function $\textrm{diag}(.)$ returns the diagonal elements of the input matrix; and $\textrm{Diag}(.)$ and $\text{Blkdiag}\left(\cdot\right)$ produce a diagonal/block-diagonal matrix with the same diagonal entries/blocks as their vector/matrices argument. The minimum eigenvalue of $\mbA$ is denoted by $\lambda_{min}(\mbA)$.
The Hadamard (element-wise) and Kronecker products are $\odot$ and $\otimes$, respectively. $l_2$-norm of  $\mba$ and Frobenius norm of 
 $\mbA$ is  denoted by $\|\mba\|_{_2}$ and $\|\mbA\|_{_{\text{F}}}$, respectively. 
 $\mathrm{vec}_{_{M,N}}^{-1}\left(\mba\right)$ reshapes the input vector $\mba\in\mathbb{C}^{MN \times 1}$ into a  matrix $\mbA\in\mathbb{C}^{M \times N}$ such that $\vec{\mbA}=\mba$.
 \vspace{-10pt}
\section{System Model}\label{sec:system-mdoel}
We consider a  network of  $K$  cooperative vehicles, each equipped with radars that have $N$ transmit antennas and $M$ receive antennas arranged as a uniform linear array (ULA). Each vehicle  transmits  $L$   FMCW chirps during the CPI time of  $T$, with similar bandwidth $B$ and  chirp time $T_c$. The transmit waveform at  $l-$th pulse  of one transmitter antenna is 
\begin{equation}
s(t,l)=\text{rect}\left(\frac{t-lT_c}{T}\right)e^{\j 2 \pi [f_c+\frac{B}{T}(t-lT_c)](t-lT_c)}.
\end{equation}
We denote the position of the $n-$th transmitter on  vehicle $k$ by $\mbp_{_{T,kn}}\in \realR{2 \times 1}$ and  the position of $m-$th receiver  on vehicle $i$ by $\mbp_{_{R,im}}\in \realR{2 \times 1}$.
We further assume that the target  at position $\mbp_{_t}\in \realR{2 \times 1}$
is moving with a velocity $\mbv_t \in \realR{2 \times 1}$.
The Doppler velocity  of the target with respect to vehicle $k$ is  
\begin{equation}
\nu_{_k}=\mbv_t^{\T}\mbp_{_{tk}},
\end{equation}
where $\mbp_{_{tk}}=[\sin{\theta_{_{tk}}},\cos{\theta_{_{tk}}}]^{\T}$ is the  direction vector of the target with DoA $\theta_{_{tk}}$ with  respect to vehicle $k$. 

The range from the $n$-th transmitter  on vehicle $k$ to the target is
\begin{equation}\label{eq:range_k}
R_{_k}(t,l,n)=\|\mbp_{_t}-\mbp_{_{R,im}}\|_{_2} + \nu_{_k} (t+(n-1+(l-1)N)T_c),
\end{equation}
and range from target to the m-th Rx  on vehicle $i$ is
\begin{equation}\label{eq:range_i}
R_{_i}(t,l,m)= \|\mbp_{_t}-\mbp_{_{R,im}}\|_{_2} + \nu_{_i} (t+(m-1+(l-1)M)T_c).
\end{equation}
Consequently, the delay introduced  from the $n$-th transmitter  on vehicle $k$ into the signal received  at the $m$-th receiver   on vehicle $i$ is
\begin{equation}\label{eq:delay}
\tau_{_{ki}}=\frac{R_{_k}(t,l,n)+R_{_i}(t,l,m)}{c},
\end{equation}
where $c$ is the speed of light.

\begin{figure}[t]
\input{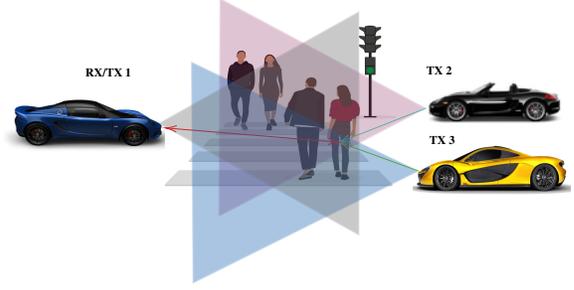}
\caption{A simplified illustration of a CAV  platoon consisting of three vehicles, sensing a target in the  FoV of all three vehicles. The radar on vehicle 1, denoted by RX/TX1, leads the platoon and is assisted by two other radars, denoted by TX2 and TX3.} 
\label{fig_1}
\end{figure}

We assume that each pair of vehicles $i,k\in \{1,\ldots,K\}$ is connected via V2V communication links. In automotive radar, the signal processing flow sequentially comprises of sampling, range estimation, Doppler processing, and DoA estimation. In STAP, after range processing the Doppler and DoA are processed simultaneously by means of 2D adaptive matched filters~\cite{bruyere2008adaptive, goodman2007optimum}. Therefore, after sampling the  signal backscattered from the target i.e.  $s(t-\tau_{_{ki}},l)$ and  estimating the  range, the received signal  at the designated range bin at the  $m$-th Rx on vehicle $i$ from the $n$-th Tx on vehicle  $k$ is 
\begin{align}
s_{_{ki}}(l,n,m)=&\alpha_k e^{-\j\frac{2\pi f_c}{c}\nu_k\left((l-1)N+(n-1)\right)T_c} \nonumber\\&
e^{-\j\frac{2\pi f_c}{c}\nu_i\left((l-1)M+(m-1)\right)T_c} \nonumber\\&
e^{\j2\pi f_c \left(\mbp_{_{T,kn}}^{\T}\mbp_{_{tk}}\right)} e^{\j2\pi f_c \left(\mbp_{_{R,im}}^{\T}\mbp_{_{ti}}\right)},
\end{align}
where the  delay in~\eqref{eq:delay} is expanded with respect to the 
first element of ULA in a similar manner as in~\cite{wang2020stap} and $\alpha_k$ is the complex target reflection factor. We introduce the  Doppler   steering  vector as
\begin{equation}
\mba_{_{d,N}}(\nu)=\left[1,e^{-\j\frac{2\pi f_c \nu}{c}N T_c},\ldots,e^{-\j\frac{2\pi f_c \nu}{c}(L-1)N T_c}\right]^{\T},
\end{equation} 
and the auxiliary Doppler steering  vector as 
\begin{equation}
\mba_{_{D,M}}(\nu)=\left[1,e^{-\j\frac{2\pi f_c \nu}{c}T_c},\ldots,e^{-\j\frac{2\pi f_c \nu}{c}(M-1) T_c}\right]^{\T}.
\end{equation} 
The transmit array steering vector at  vehicle $k$ is
\begin{equation}
\mba_{_{T,k}}(\theta)=\left[e^{\j 2\pi f_c\left(\mbp_{_{T,k1}}^{\T}\mbp_{_{tk}}\right)},\ldots,e^{\j 2\pi f_c \left(\mbp_{_{T,kN}}^{\T}\mbp_{_{tk}}\right)}\right]^{\T},
\end{equation}
and the receive array steering vector at vehicle $i$ is 
\begin{equation}
\mba_{_{R,i}}(\theta)=\left[e^{\j2\pi f_c \left(\mbp_{_{R,i1}}^{\T}\mbp_{_{ti}}\right)},\ldots,e^{\j2\pi f_c\left(\mbp_{_{R,iM}}^{\T}\mbp_{_{ti}}\right)}\right]^{\T}.
\end{equation}
The  snapshot signal received at  vehicle $i$  from $n$-th Tx on vehicle $k$ is 
\begin{align}\label{eq:s_ki}
\mbs_{_{ki}}(n)= & \left[\mba_{_{d,N}}(\nu_k)\odot \mba_{_{d,M}}(\nu_i) \right] \nonumber\\
&\otimes \left[\mba_{_{R,i}}(\theta) \odot \mba_{_{D,M}}(\nu_i)\right] \in \complexC{LM \times 1}.
\end{align}
By stacking the   echoes from all N Tx  on vehicle $k$, we obtain
\begin{align}\label{s_kin}
\mbs_{_{ki}}&=\begin{bmatrix}
\mbs_{_{ki}}(1)\\
 \vdots\\
\mbs_{_{ki}}(N)\\
\end{bmatrix}\nonumber \\
&= \left(\mba_{_{T,k}}(\theta) \odot \mba_{_{D,N}}(\nu_k) \right)\otimes    \Big(\left(\mba_{_{d,N}}(\nu_k)\odot \mba_{_{d,M}}(\nu_i) \right)
\nonumber \\
& \;\;\otimes  \left(\mba_{_{R,i}}(\theta) \odot \mba_{_{D,M}}(\nu_i)\right) \Big) \in \complexC{NLM \times 1}.
\end{align}
 As illustrated in Fig.~\ref{fig_1}, we focus on a scenario wherein vehicle 
$i$, serving as the lead in the platoon, receives assistance from all other vehicles for its sensing task. The target is in the field of view (FoV) of all vehicles and therefore the whole CAV can perform as a multistatic radar to sense it.  To be   succinct, hereafter we  remove the  subscript $i$ i.e.  without  loss of generality $\mbs_{ki}$ will be replaced by $\mbs_{k}$.  
In order  to decide whether a 
target is present in a particular known range-cell, we perform binary hypothesis testing between $\mathcal{H}_0$ (target-free hypothesis) and $\mathcal{H}_1$ (target-present hypothesis), i.e., 
\begin{align}
\calH_0 &:\quad \mby_k= \mbn_k  \nonumber\\
\calH_1 &: \quad \mby_k=\alpha_k\mbs_k+\mbn_k, 
\end{align}
where $\alpha_k$ is the  complex target reflectivity factor and $\mbn_k$ is the  noise and interference with covariance $\mbR_k$~\cite{esmaeilbeig2023moving}. The log-likelihood ratio  test statistic is given by~\cite{goodman2007optimum},
\begin{equation}\label{eq::test_statistics}
\zeta=\sum_{k=1}^{K} \frac{|\mbs_k^{\H}\mbR_k^{-1}\mby_k|^2}{\frac{|\alpha_k|^{2}}{2}+\mbs_k^{\H}\mbR_k^{-1}\mbs_k} \underset{\calH_0}{\overset{\calH_1}{\gtrless}} \gamma.
\end{equation}
Since $|\mbs_k^{\H}\mbR_k^{-1}\mby_k|$ is the
magnitude of a complex Gaussian random variable,
the test statistic in~\eqref{eq::test_statistics} is a sum of $K$ exponential random variables  and  therefore  follows a  hypo-exponential distribution~\cite{rouffet2016analysis}.  

Under  $\calH_1$, we have 
\begin{equation}\label{eq:mean-H_1}
\expecE{\zeta | \calH_1}= \sum_{k=1}^{K}\frac{\mbC_k+2|\alpha_k|^2\mbC_k^2}{\frac{|\alpha_k|^2}{2}+\mbC_k}, \end{equation}
where $\mbC_k=\mbs_k^{\H}\mbR_k^{-1}\mbs_k$. The probability of  detection  and false alarm are  obtained  respectively as 
\begin{align}
\text{P}_{_{\text{D}}}&=\probP{\zeta > \gamma | \calH_1}=1-\probP{\zeta \leq \gamma | \calH_1}=1- F_{\zeta|\calH_1}(\gamma | \calH_1),\nonumber\\
\text{P}_{_{\text{FA}}}&=\probP{\zeta > \gamma | \calH_0}=1-\probP{\zeta \leq \gamma | \calH_0}=1- F_{\zeta|\calH_0}(\gamma | \calH_0),
\end{align}
where $F_{\zeta|\calH}(.)$ is the cumulative  distribution  function of the test statistic with  hypo-exponential  distribution. By carefully observing~\eqref{eq::test_statistics}, one can verify that the weighting $(\frac{|\alpha_k|^{2}}{2}+\mbs_k^{\H}\mbR_k^{-1}\mbs_k)^{-1}$ accounts for the contribution of  vehicle $k$ in the  test statistic. If $\mbs_k^{\H}\mbR_k^{-1}\mbs_k \ll \frac{|\alpha_k|^{2}}{2}$  then  we can interpret it as  the signal propagated from  vehicle $k$  not  reaching the receiver. Therefore, for the  sake of interpretation  here we assume $\mbs_k^{\H}\mbR_k^{-1}\mbs_k \gg \frac{|\alpha_k|^{2}}{2}$ for all  vehicles i.e.  $k\in \{1,\ldots,K\}$. Under this  condition,~\eqref{eq:mean-H_1} is
\begin{equation}\label{eq:mean}
\expecE{\zeta | \calH_1} = \sum_{k=1}^{K} 1+2|\alpha_k|^2\mbC_k
\end{equation}
The signal received at  vehicle $i$ from all vehicles, i.e.,  $k \in \{1,\ldots,K\}$, is thus given by $\mbs=[\mbs_{_{1}}, \cdots, \mbs_{_{K}}]^T \in  \complexC{KNLM}$, 
 where $\mbs$ is also equivalent to the  2D space-time steering  vector  of the target as observed at vehicle  $i$.
\section{TDM design}\label{sec:TDM}
Since multiple signals share the same communication channel, the transmitted signals need to be orthogonal to be distinguishable at the receiver.  For this matter in a CPI  the antennas need to take turns to transmit. We propose to incorporate  TDM by designing a transmitter scheduling matrix for the platoon of  vehicles.  In this particular TDM scheme, at most one antenna within the platoon is allowed to transmit during each pulse. We reformulate the  steering vectors  under the TDM scheme. If  $\mbe_{_{kn}} $ is a one-hot  vector  of size $L \times 1 $ with only one element as unity and remaining elements zero, then under TDM, ~\eqref{eq:s_ki}  is 
\begin{equation}
\mbs_{_{ki}}(n)=\left[\mbe_{_{kn}}\odot \mba_{_{d,N}}(\nu_k)\odot \mba_{_{d,M}}(\nu_i) \right] \otimes \left[\mba_{_{R,i}}(\theta) \odot \mba_{_{D,M}}(\nu_i)\right],    
\end{equation}
where the unity element in  $\mbe_{kn}$, indicates  the  pulse  at which the  transmitter antenna $n$ on vehicle  $k$  transmits. Let $\mbJ_k=\left[\mbe_{_{k1}}\bigm\lvert\ldots \bigm\lvert\mbe_{_{kN}}\right]\in\{0,1\}^{L \times N}$. Following  the same procedure leading to~\eqref{s_kin}, we have
\begin{align}
\bar{\mbs}_{_{k}}
&=\Big(\Vec{\mbJ_k}\otimes \bone  \Big) \odot \Big( \left(\mba_{_{T,k}}(\theta) \odot \mba_{_{D,N}}(\nu_k) \right)\otimes  \nonumber \\   & \left(\mba_{_{d,N}}(\nu_k)\odot \mba_{_{d,M}}(\nu_i) \right)
\otimes \left(\mba_{_{R,i}}(\theta) \odot \mba_{_{D,M}}(\nu_i)\right) \Big) \nonumber\\
&=\Big(\Vec{\mbJ_k}\otimes \bone  \Big) \odot  \mbs_k
\end{align}
After performing TDM the received signal  $\bar{\mbs}=[\bar{\mbs}_{_{1}}, \cdots, \bar{\mbs}_{_{K}}]^T$ is 
\begin{align}\label{eq:overall-steering2}
\bar{\mbs}&=\begin{bmatrix}
\Big(\Vec{\mbJ_1}\otimes \bone  \Big) \odot\mbs_{_{1}}\\
 \vdots\\
\Big(\Vec{\mbJ_k}\otimes \bone  \Big) \odot\mbs_{_{K}}\\
\end{bmatrix}=\Big(\vec{[\mbJ_1 \bigm\lvert \ldots \bigm\lvert \mbJ_K]}\otimes \bone_{M} \Big) \odot \mbs \nonumber\\&=\Big(\vec{\mbJ}\otimes \bone_{M} \Big) \odot \mbs\in\complexC{KNLM\times 1},
\end{align}

where $\mbJ=\left[\mbJ_1 \bigm\lvert \ldots \bigm\lvert \mbJ_K\right] \in \{0,1\}^{L\times KN}$ is the waveform selection matrix. Without loss of  generality, we design a   TDM  scheme in which the  antennas in the whole platoon take turns over the pulses to  transmit. Under this scheme   for  a CPI of  length  $L=KN$,  each antenna selected to transmit at pulse  $l$, prevents all other  antennas from transmitting. As a  consequence, the  waveform  selection matrix  $\mbJ$ is  a permutation matrix of    size 
 $L\times L$.  
We intend to maximize the target detection performance. We use the mean of the test statistic as the design criteria. Consequently the  TDM design problem is  
\begin{align}
\label{eq:opt1}
\mathcal{P}_{1}:\;\underset{\mbJ}{\text{maximize}} &\quad 
\expecE{\zeta | \calH_1}
 \nonumber\\
\text{subject to} &\quad 
\sum_{p}\mbJ_{pn}=1,  \quad \quad p,n \in \{1,\ldots,L\};
\nonumber\\
&\quad \sum_{n}\mbJ_{pn}=1 
;\nonumber\\
&\quad ~~\mbJ_{pn}\in \{0,1 \}. 
\end{align}
\section{solution methodology}\label{sec:method}
In this section, we  first  demonstrate that  $\mathcal{P}_{1}$ is a QAP~\cite{burkard1998quadratic}
which is a  combinatorial optimization problem that is   NP-hard in general form. 
Consequently, we introduce a computationally efficient procedure to obtain a local optimum of QAP. Our proposed method takes advantage of the power method-like iterations introduced in~\cite{soltanalian2014designing}, which resembles the well-known power method for computing the dominant eigenvalue and vector pairs of matrices.

We accumulate the interference  covariance matrices in $\mbR$ such that $\mbR=\text{Blkdiag}\left(\mbR_1,\ldots,\mbR_K\right)$. By substituting~\eqref{eq:overall-steering2} in~\eqref{eq:mean} we  obtain\par \noindent\small
\begin{align}
 &\expecE{\zeta| \calH_1}=\bar{\mbs}^{\H}\mbR^{-1}\bar{\mbs}\nonumber\\
 &=\left(\Big(\vec{\mbJ}\otimes \bone_{M} \Big) \odot \mbs\right)^{\H} \mbR^{-1}\left(\Big(\vec{\mbJ}\otimes \bone_{M} \Big) \odot \mbs\right)\nonumber\\
 &=\Big(\vec{\mbJ}\otimes \bone_{M}\Big)^{\H}\Diag{\mbs}^{\H}\mbR^{-1} 
 \Diag{\mbs}\Big(\vec{\mbJ}\otimes \bone_{M}\Big)\nonumber\\
 &= \Tr{\Big(\vec{\mbJ}\otimes \bone_{M}\Big)^{\H}\mbQ\Big(\vec{\mbJ}\otimes \bone_{M}\Big)}\nonumber\\
 &=\Big(\vec{\mbJ}\otimes \bone_{M}\Big)^{\H}\vec{\mbQ\Big(\vec{\mbJ}\otimes \bone_{M}\Big)} \nonumber \\
 &= \vec{\mbJ}^{\H}\mbG^{\H}\mbQ \;\mbG \; \vec{\mbJ}
 \end{align}\normalsize
 where 
 \begin{align}
 \mbQ&=\Diag{\mbs}^{\H} \mbR^{-1} \Diag{\mbs}\\
 \mbG&=(\mbK_{_{1,KNL}}\otimes \mbI_M)(\mbI_{KNL}\otimes \bone_{M})
 \end{align}
 and $\mbK_{_{1,KNL}}$ is the commutation matrix satisfying
 \begin{eqnarray}
     \mbK_{_{1,KNL}}\vec{\mbJ}=\vec{\mbJ^{\T}}.
 \end{eqnarray}
 The above algebraic manipulations cast $\mathcal{P}_{1}$  equivalent to a  QAP as
\begin{align}
\label{eq:opt2}
\mathcal{P}_{2}:~\underset{\mbJ \in \bOmega}{\textrm{maximize}} &\quad \vec{\mbJ}^{\H} \mbS \;\vec{\mbJ},
\end{align}
where $\mbS=\mbG^{\H}\mbQ \mbG$  and $\bOmega$ is the set of permutation  matrices  i.e.  
\begin{align}
\bOmega=\Bigg \{\mbJ\; \biggm\lvert \;\sum_{p}\mbJ_{pn}=1,&\quad \sum_{n}\mbJ_{pn}=1,\quad\\
&\mbJ_{pn}\in\{0,1\}, \quad p,n \in \{1,\ldots,L\}\Bigg \}.   \nonumber 
\end{align}
 By performing  diagonal loading, i.e.,  substituting   $\mbS$  with positive semi-definite matrix $\bar{\mbS}=\lambda_m \mbI+\mbS$, with $\lambda_m \geq -\lambda_{min}(\mbS)$, $\mathcal{P}_{2}$ will be  turned into  the equivalent  maximization problem:
\begin{align}
\label{eq:opt3}
\mathcal{P}_{3}:~\underset{\mbJ \in \bOmega}{\textrm{maximize}} &\quad \vec{\mbJ}^{\H}\bar{\mbS} \; \vec{\mbJ}.
\end{align}
One can locally optimize $\mathcal{P}_{3}$ by resorting to \textit{power method-like} iterations of the  form \cite{soltanalian2014designing,soltanalian2013joint}: 
\begin{align}
\label{eq:opt4}
\mathcal{P}_{4} &:
~\underset{\mbJ^{(s+1)}\in \bOmega}{\textrm{minimize}}
\quad \|\vec{\mbJ^{(s+1)}}-\bar{\mbS}\;\vec{\mbJ^{(s)}}\|_{_{2}} \nonumber\\ 
\equiv&\quad\underset{\mbJ^{(s+1)}\in \bOmega}{\textrm{minimize}}
\quad \|\mbJ^{(s+1)}-\mathrm{vec}_{_{L,L}}^{-1}\left(
\bar{\mbS}\;\vec{\mbJ^{(s)}}\right)\|_{_{\text{F}}}.
\end{align}
We define  the  matrix $\mbC^{(s)}=-\mathrm{vec}_{_{L,L}}^{-1}\left(\bar{\mbS} \;\vec{\mbJ^{(s)}}\right)$. It is  straightforward to see \par \noindent \small
\begin{align}
& \|\mbJ^{(s+1)}+\mbC^{(s)}\|^2_{_{\text{F}}}=\Tr{(\mbJ^{(s+1)}+\mbC^{(s)})^{\H}(\mbJ^{(s+1)}+\mbC^{(s)})}\nonumber\\
&=\Tr{\mbI+\mbC^{(s)\H}\mbC^{(s)}}+\Tr{\mbC^{(s)\H}\mbJ^{(s+1)}+\mbJ^{(s+1)\H}\mbC^{(s)}}\nonumber\\
&= \Tr{\mbI+\mbC^{(s)\H}\mbC^{(s)}}+2\Tr{\mbJ^{(s+1)}\mbC^{(s)\H}},
\end{align}\normalsize
where we used  the orthogonality property of  permutation matrices i.e. $\mbJ^{(s+1)\H}\mbJ^{(s+1)}=\mbI$ in the second equality. Consequently, $\mathcal{P}_{4}$ is equivalent to
\begin{align}
\label{eq:opt5}
\mathcal{P}_{5}:& \quad \underset{\mbJ^{(s+1)}\in \bOmega}{\textrm{minimize}}
\quad  \Tr{\mbJ^{(s+1)}\mbC^{(s)\H}}.
\end{align}
Note that the above problem is in fact a \textit{linear assignment problem} with cost matrix $\mbC^{(s)\H}$ that can be solved  efficiently  using the  \textit{Hungarian algorithm} also known as  \textit{Munkres assignment algorithm}, with  computational complexity of $\mathcal{O}(L^2)$~\cite{kuhn1955hungarian}. Our final proposed  algorithm  for transmitter scheduling in  CAVs based on power method-like iterations is presented in Algorithm 1. As shown in~\cite{soltanalian2014designing}, the  objective $f(\mbJ)= \vec{\mbJ}^{\H}\bar{\mbS} \; \vec{\mbJ}$ is increasing through the power method-like iterations and  convergent in
the sense of the objective value. Consequently, we set the stopping criteria $\bigm| \bigm[f(\mbJ^{(s+1)})-f(\mbJ^{(s)})\bigm]/f(\mbJ^{(s)})\bigm| < \epsilon$ for the  algorithm.\vspace{-6pt}
\begin{algorithm}[t]
\caption{Power method-like iterations for  transmitter scheduling in  CAVs.}
    \label{alg:place}
    \begin{algorithmic}[1]
    \State \textbf{Input} The overall steering vector of the CAV $\bar{\mbs}$
    \State \textbf{Initialization}  $\mbJ^{(0)} \in \bOmega$, $s=0$
    \State $\bar{\mbS}=\lambda_m \mbI-\mbS$
    \State  \textbf{While} $\bigm|\bigm[f(\mbJ^{(s+1)})-f(\mbJ^{(s)})\bigm]/f(\mbJ^{(s)})\bigm|\geq \epsilon$ \textbf{do} 
    \State  \hspace{1cm}$\mbC^{(s)}=-\mathrm{vec}_{_{L,L}}^{-1}\left(\bar{\mbS} \;\vec{\mbJ^{(s)}}\right)$
    \State \hspace{1cm} $\mbJ^{(s+1)}\leftarrow  \text{Hungarian}(\mbC^{(s)\H})$
    \State \hspace{1cm} $s\leftarrow  s+1$
    \State $\mbJ_{_{\text{opt}}}\leftarrow \mbJ^{(s)}$
    \State  \textbf{Output} $\mbJ_{_{\text{opt}}}$
   \end{algorithmic}
\end{algorithm}
\section{Numerical Experiments}\label{sec_5}
\begin{figure}[t]
\vspace{-.75cm} 
\hspace{-.35cm}
\includegraphics[width=1.1\columnwidth]{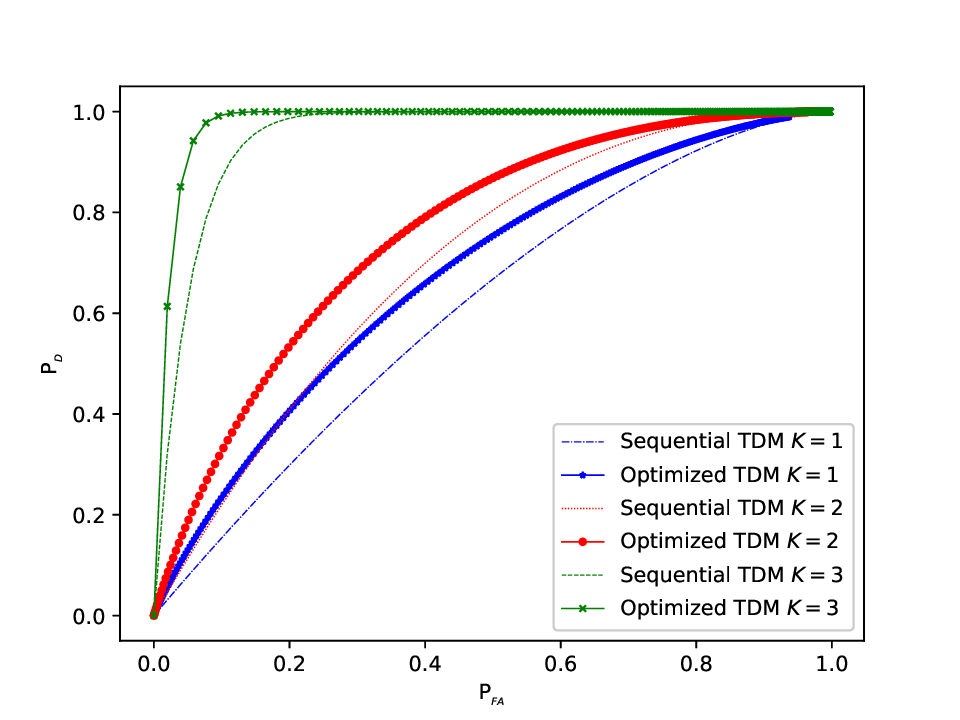}
\caption{RoC of detection for a CAV of FMCW radars. The optimized  TDM  is compared  with uniform transmission where the  antennas are  activated uniformly in a  sequence.}
\label{fig::2}
\end{figure}

We carried out numerical  experimentation
to evaluate the performance of the proposed algorithm for TDM in CAVs. We considered  a platoon consisting of $K=3$ vehicles, each equipped with $M=N=8$ Tx and Rx antennas arranged as ULA. All vehicles are assumed to have FMCW radar  systems operating at carrier  frequency $f_c=77$ GHz, bandwidth  $B=150$ MHz and  chirp time $T_c=8 \; \mu$s. The  vehicles are moving with  2D velocities  $\mbv_1=[20,20]$ m/s, $\mbv_2=[-10,-20]$ m/s and  $\mbv_3=[30,15]$ m/s. Algorithm 1 is performed to  design the  TDM for a  CPI of   length $L=24$ pulses. The receiver operating characteristics (RoC) associated with the described CAV when one, two or three of the vehicles are  cooperating and  transmitting  with optimized TDM  is illustrated in Fig.~\ref{fig::2}. The results  demonstrate  significant improvement  in  comparison with sequential transmission   where  the transmitters on each vehicle are activated by the order of their indices $\{1,2,\ldots,N\}$, equivalent to an identity TDM matrix.
\section{Summary}\label{sec:conclusion}
Vehicle-to-vehicle communication is a fundamental capability that enables CAVs to perform distributed STAP. A  cooperative  STAP scheme for the vehicles in a CAV platoon was within the purview of  this paper.  Furthermore, we  introduced a TDM  scheme for   orthogonal  transmission of the antennas  in the CAV platoon. The transmitter  scheduling was   formulated as a  quadratic assignment problem.  By means of the  well-known  power method-like iterations we showed that the TDM design problem can be  reduced to  a linear assignment  problem in each iteration and  consequently the efficient Hungarian algorithm can effectively address it. Through numerical simulations, we confirmed the efficacy of the suggested TDM approach in enhancing target detection performance.
\bibliographystyle{IEEEtran}
\bibliography{refs}

\begin{thebibliography}{10}
\providecommand{\url}[1]{#1}
\csname url@samestyle\endcsname
\providecommand{\newblock}{\relax}
\providecommand{\bibinfo}[2]{#2}
\providecommand{\BIBentrySTDinterwordspacing}{\spaceskip=0pt\relax}
\providecommand{\BIBentryALTinterwordstretchfactor}{4}
\providecommand{\BIBentryALTinterwordspacing}{\spaceskip=\fontdimen2\font plus
\BIBentryALTinterwordstretchfactor\fontdimen3\font minus
  \fontdimen4\font\relax}
\providecommand{\BIBforeignlanguage}[2]{{%
\expandafter\ifx\csname l@#1\endcsname\relax
\typeout{** WARNING: IEEEtran.bst: No hyphenation pattern has been}%
\typeout{** loaded for the language `#1'. Using the pattern for}%
\typeout{** the default language instead.}%
\else
\language=\csname l@#1\endcsname
\fi
#2}}
\providecommand{\BIBdecl}{\relax}
\BIBdecl

\bibitem{sarker2020}
A.~Sarker, H.~Shen, M.~Rahman, M.~Chowdhury, K.~Dey, F.~Li, Y.~Wang, and H.~S.
  Narman, ``A review of sensing and communication, human factors, and
  controller aspects for information-aware connected and automated vehicles,''
  \emph{IEEE Transactions on Intelligent Transportation Systems}, vol.~21,
  no.~1, pp. 7--29, 2020.

\bibitem{wang2023anomaly}
Y.~Wang, R.~Zhang, N.~Masoud, and H.~X. Liu, ``Anomaly detection and string
  stability analysis in connected automated vehicular platoons,''
  \emph{Transportation research part C: emerging technologies}, vol. 151, p.
  104114, 2023.

\bibitem{Griffiths}
H.~Griffiths, ``Multistatic, {MIMO} and networked radar: The future of radar
  sensors?'' pp. 81--84, 2010.

\bibitem{zhangjoint2022}
H.~Zhang, W.~Liu, J.~Shi, T.~Fei, and B.~Zong, ``Joint detection threshold
  optimization and illumination time allocation strategy for cognitive tracking
  in a networked radar system,'' \emph{IEEE Transactions on Signal Processing},
  vol.~70, pp. 5833--5847, 2022.

\bibitem{goodman2007optimum}
N.~A. Goodman and D.~Bruyere, ``Optimum and decentralized detection for
  multistatic airborne radar,'' \emph{IEEE Transactions on Aerospace and
  Electronic Systems}, vol.~43, no.~2, pp. 806--813, 2007.

\bibitem{radarsignaldesign2022}
M.~Alaee-Kerahroodi, M.~Soltanalian, P.~Babu, and M.~R.~B. Shankar,
  \emph{Signal Design for Modern Radar Systems}.\hskip 1em plus 0.5em minus
  0.4em\relax Artech House, 2022.

\bibitem{wang2020stap}
G.~Wang, K.~V. Mishra \emph{et~al.}, ``{STAP} in automotive {MIMO} radar with
  transmitter scheduling,'' in \emph{IEEE Radar Conference}, 2020, pp. 1--6.

\bibitem{bose2021mutual}
A.~Bose, B.~Tang, M.~Soltanalian, and J.~Li, ``Mutual interference mitigation
  for multiple connected automotive radar systems,'' \emph{IEEE Transactions on
  Vehicular Technology}, vol.~70, no.~10, pp. 11\,062--11\,066, 2021.

\bibitem{tang2016joint}
B.~Tang and J.~Tang, ``Joint design of transmit waveforms and receive filters
  for {MIMO} radar space-time adaptive processing,'' \emph{IEEE Transactions on
  Signal Processing}, vol.~64, no.~18, pp. 4707--4722, 2016.

\bibitem{tang2020polyphase}
B.~Tang, J.~Tuck, and P.~Stoica, ``Polyphase waveform design for {MIMO} radar
  space time adaptive processing,'' \emph{IEEE Transactions on Signal
  Processing}, vol.~68, pp. 2143--2154, 2020.

\bibitem{bose2022waveform}
A.~Bose, B.~Tang, W.~Huang, M.~Soltanalian, and J.~Li, ``Waveform design for
  mutual interference mitigation in automotive radar,'' \emph{arXiv preprint
  arXiv:2208.04398}, 2022.

\bibitem{esmaeilbeig2023mutual}
Z.~Esmaeilbeig, A.~Bose, and M.~Soltanalian, ``Mutual interference mitigation
  in {PMCW} automotive radar,'' 2023.

\bibitem{xu2023automotive}
L.~Xu, S.~Sun, K.~V. Mishra, and Y.~D. Zhang, ``Automotive {FMCW} radar with
  difference co-chirps,'' \emph{IEEE Transactions on Aerospace and Electronic
  Systems}, pp. 1--19, 2023.

\bibitem{kuhn1955hungarian}
H.~W. Kuhn, ``The hungarian method for the assignment problem,'' \emph{Naval
  research logistics quarterly}, vol.~2, no. 1-2, pp. 83--97, 1955.

\bibitem{bose2017non}
A.~Bose and M.~Soltanalian, ``Non-convex shredded signal reconstruction via
  sparsity enhancement,'' in \emph{IEEE International Conference on Acoustics,
  Speech and Signal Processing}, 2017, pp. 4691--4695.

\bibitem{bruyere2008adaptive}
D.~P. Bruyere and N.~A. Goodman, ``Adaptive detection and diversity order in
  multistatic radar,'' \emph{IEEE Transactions on Aerospace and Electronic
  Systems}, vol.~44, no.~4, pp. 1615--1623, 2008.

\bibitem{esmaeilbeig2023moving}
Z.~Esmaeilbeig, A.~Eamaz, K.~V. Mishra, and M.~Soltanalian, ``Moving target
  detection via multi-{IRS}-aided {OFDM} radar,'' in \emph{IEEE Radar
  Conference}.\hskip 1em plus 0.5em minus 0.4em\relax IEEE, 2023, pp. 1--6.

\bibitem{rouffet2016analysis}
T.~Rouffet, P.~Vallet, E.~Grivel, C.~Enderli, B.~Joseph, and S.~Kemkemiant,
  ``Analysis of a {GLRT} for the detection of an extended target,'' in
  \emph{IEEE Radar Conference}.\hskip 1em plus 0.5em minus 0.4em\relax IEEE,
  2016, pp. 1--5.

\bibitem{burkard1998quadratic}
R.~E. Burkard, E.~Cela, P.~M. Pardalos, and L.~S. Pitsoulis, \emph{The
  quadratic assignment problem}.\hskip 1em plus 0.5em minus 0.4em\relax
  Springer, 1998.

\bibitem{soltanalian2014designing}
M.~Soltanalian and P.~Stoica, ``Designing unimodular codes via quadratic
  optimization,'' \emph{IEEE Transactions on Signal Processing}, vol.~62,
  no.~5, pp. 1221--1234, 2014.

\bibitem{soltanalian2013joint}
M.~Soltanalian, B.~Tang, J.~Li, and P.~Stoica, ``Joint design of the receive
  filter and transmit sequence for active sensing,'' \emph{IEEE Signal
  Processing Letters}, vol.~20, no.~5, pp. 423--426, 2013.

\end{thebibliography}
\end{document}